\documentclass[aps,10pt,twocolumn,amsmath,amssymb,prl,superscriptaddress]{revtex4-1}
\usepackage{graphicx}
\usepackage{dcolumn}
\usepackage{bm}

\setcitestyle{super}

\begin{document}

\title{Tunable non-equilibrium Luttinger liquid based on counter-propagating edge channels}
\author{M.G.~Prokudina}
\affiliation{Institute of Solid State Physics, Russian Academy of
Sciences, 142432 Chernogolovka, Russian Federation}
\author{S.~Ludwig}
\affiliation{Center for NanoScience and Fakult\"{a}t f\"{u}r Physik,
Ludwig-Maximilians-Universit$\ddot{\text{a}}$t,
Geschwister-Scholl-Platz 1, D-80539 M$\ddot{\text{u}}$nchen,
Germany}
\author{V. Pellegrini}
\affiliation{Istituto Italiano di Tecnologia (IIT), Via Morego 30, 16163 Genova, Italy
NEST, Istituto Nanoscienze-CNR and Scuola Normale Superiore, I-56126 Pisa, Italy}
\author{L. Sorba}
\affiliation{NEST, Istituto Nanoscienze-CNR and Scuola Normale Superiore, Piazza San Silvestro 12, I-56127 Pisa, Italy}
\author{G.~Biasiol}
\affiliation{CNR-IOM, Laboratorio TASC, Area Science Park, I-34149 Trieste, Italy }
\author{V.S.~Khrapai}
\affiliation{Institute of Solid State Physics, Russian Academy of
Sciences, 142432 Chernogolovka, Russian Federation}
\affiliation{Moscow Institute of Physics and Technology, Dolgoprudny, 141700 Russian Federation}

\begin{abstract}

We investigate energy transfer between counter-propagating quantum Hall edge channels (ECs) in a two-dimensional electron system at filling factor $\nu=1$. The ECs are separated by a thin impenetrable potential barrier and Coulomb coupled, thereby constituting a quasi one-dimensional analogue of a spinless Luttinger liquid (LL). We drive one, say hot, EC far from thermal equilibrium and measure the energy transfer rate $P$ into the second, cold, EC using a quantum point contact as a bolometer. The dependence of $P$ on the drive bias indicates breakdown of the momentum conservation, whereas $P$ is almost independent on the length of the region where the ECs interact. Interpreting our results in terms of plasmons (collective density excitations), we find that the energy transfer between the ECs occurs via plasmon backscattering at the boundaries of the LL. The backscattering probability is determined by the LL interaction parameter and can be tuned by changing the width of the electrostatic potential barrier between the ECs.

\end{abstract}

\maketitle

One-dimensional electronic systems (1DESs) are collective in nature. As first shown by Tomonaga~\cite{tomonaga}, an interacting 1DES near its ground state can be modeled with the help of a  bosonization technique. Later on Luttinger~\cite{luttinger} introduced an exactly soluble~\cite{LiebMattis} model for two species of fermions (left and right movers) with an infinite linear dispersion, referred to as a Luttinger liquid (LL). The excitations of a spinless LL can be described as non-interacting plasmons, bosonic collective fluctuations of the electron density~\cite{giamarchi}. The plasmon's lack of interaction gives rise to the counterintuitive prediction that an excited ideal LL should never thermalize.

The strength of the LL model fully manifests itself out of equilibrium, where it still offers a single-particle description of the kinetics of a strongly correlated 1DES. In the presence of disorder, energy relaxation is then described as elastic plasmon scattering off inhomogeneities~\cite{gutmanPRB2009,bagretsPRB2009}. This has not yet been confirmed experimentally.  Instead, for two tunnel-coupled quantum wires far from equilibrium deviations from the LL model were observed~\cite{tunnel_yacoby}. The main problem seems to be disorder, which gives rise to thermalization on the length scale of the mean free path~\cite{bagretsPRB2009}. Signatures of disordered LLs, such as a powerlaw dependence of the conductance on temperature or bias, have been observed in various 1DESs \cite{bockrath,auslaender,segovia}. However, the design of experiments far from equilibrium remains complicated because of small mean free paths of no more than a few micrometers in such devices~\cite{RMP_Charlier}.

Here we apply a strong magnetic field perpendicular to the 2DES of a GaAs/AlGaAs heterostructure and realize a tunable LL based on ECs at integer filling factor $\nu=1$. Related to their chiral nature, ECs offer the fundamental advantage of suppressed back-scattering of electrons \cite{buettiker}. Yet, contrary to the chiral-LLs in fractional quantum Hall regime~\cite{grayson_refs1}, a single EC at $\nu=1$ behaves as a perfect one-dimensional Fermi liquid~\cite{FisherGlazman,hilke2001}. To still create a spinless LL we bring two counter-propagating ECs into interaction, providing left and right movers according to the original proposal by Luttinger. Here we follow Ref.~\cite{Oreg_Finkelstein}, where a direct analogy between such a system and the LL model has been demonstrated. Unlike in experiments on tunneling in cleaved edge overgrown~\cite{kang} and corner-overgrown~\cite{grayson_refs2} structures, we block the charge current between the ECs and study the energy transfer between the left and right movers in this hand-made LL. Besides  much weaker disorder this system has a second important advantage, namely the possibility of individual control over left versus right movers. 

\begin{figure}[t]
 \begin{center}
  \includegraphics[width=\columnwidth]{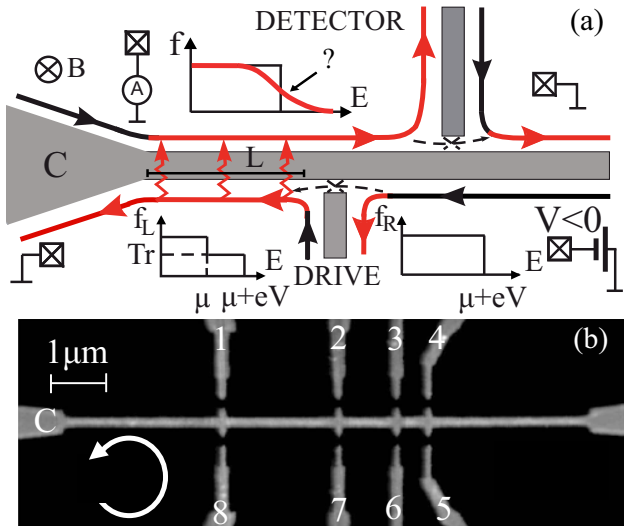}
   \end{center}
  \caption{Experimental layout. (a) -- Gated areas of the 2DES are shown in grey and the ECs are shown by solid lines with arrows. In the drive circuit we create a nonequilibrium  particle distribution in a hot EC by use of a partially transparent drive QPC biased with a voltage $V_{DRIVE}$. The hot EC propagates along the central barrier (C), reaches the interaction region of length $L$ and heats a counter-propagating cold EC in the detector circuit (winding arrows). The nonequilibrium distribution in the cold EC is characterized with the help of the detector QPC. (b) -- Electron micrograph of the sample identical to the one used in experiment. The central gate (C) and a number of side gates used to define constrictions have a grey color. The EC chirality is the same as in (a), see white arrow.}\label{fig1}
\end{figure}

Our counter-propagating ECs are separated by a barrier impenetrable for electrons, marked by C in Fig.~\ref{fig1}a. It is created electrostatically by applying a negative voltage $V_C$ to the metallic center gate (C in Fig.~\ref{fig1}b). Varying $V_C$ allows to tune the width of the barrier and the strength of the Coulomb coupling between the ECs. Other gates (1 through 8 in Fig.~\ref{fig1}b) have two purposes: first, they can be used to control the length of the interaction region ($L$) by guiding the ECs away from the center barrier. Their second purpose is to define QPCs. We create a nonequilibrium electronic distribution in one, say hot, EC by partitioning the electrons at a drive QPC~\cite{kouwenhoven,altimiras} (DRIVE circuit in Fig.~\ref{fig1}a). Based on this technique, energy relaxation between co-propagating ECs was already investigated at $\nu=2$ with a quantum dot as detector \cite{altimiras,altimiras_natphys2010} and in the fractional quantum Hall regime by observation of complex edge reconstruction effects \cite{yacoby}. We use a second QPC, defined in the counter-propagating EC (DETECTOR circuit in Fig.~\ref{fig1}a), to detect the excess energy transferred from the hot EC. This setup allows us to create a perfect LL model system out of equilibrium and to study the energy flux between the left and right movers (red winding arrows in Fig.~\ref{fig1}a).

Our samples are based on a 200~nm deep 2DES of a GaAs/AlGaAs heterostructure with electron density $9.3\times10^{10}$~cm$^{-2}$ and mobility 4$\times10^6$~cm$^2$/Vs. The metallic gates are obtained by thermal evaporation of 3~nm Ti and 30~nm Au. The experiments are performed in a $^3$He/$^4$He dilution refrigerator in a magnetic field of 3.8~T at 60 and 90~mK. Current measurements are performed using home-made $I-V$ converters with input offset voltage $\leq10\mu$V. In bolometric experiments, the detector QPC conductance is measured with a $5\mu$V rms ac modulation (11-33~Hz). For thermoelectric measurements we use a fixed ac modulation (11-33~Hz)  and measure the derivative $dI_{DET}/dV_{DRIVE}$ as a function of the dc bias $V_{DRIVE}$, which is numerically integrated to give $I_{DET}$. The modulation is $5\mu$V rms at $|V_{DRIVE}|\leq300\mu$V, and $30\mu$V rms otherwise. Throughout the paper the drive QPCs conductance is $\approx0.3e^2/h$, corresponding to a perfectly linear $I$-$V$. Hence, the excitation is the same for both polarities of $V_{DRIVE}$, explaining almost perfect symmetry of the data in figs.~\ref{fig2} and~\ref{fig3} below.

\begin{figure}[t]
 \begin{center}
  \includegraphics[width=\columnwidth]{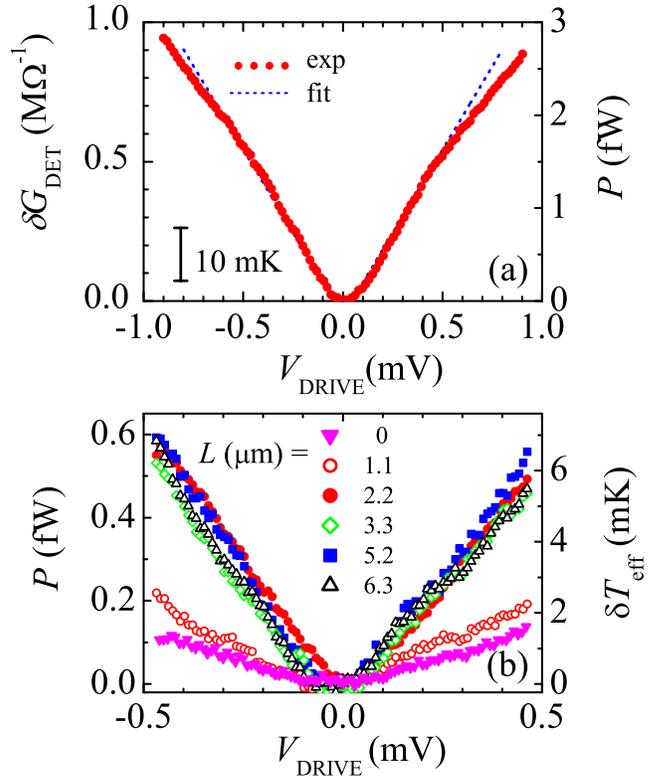}
   \end{center}
\caption{QPC as a bolometer. (a) -- Bolometric response $\delta G_{DET}$ versus  $V_{DRIVE}$ (dots). The detector QPC is defined by the gate 2 and the drive QPC with the gate 5, $L=5.2\, \mu$m. The energy transfer rate $P$ is shown on right axis scale. Vertical scale bar corresponds to $\delta T_{eff}=10$~mK. The dashed line is a fit to the model of boundary plasmon scattering with parameters $\varepsilon_0=80\mu$eV, $K\approx1.12$. The data were taken at $V_C=-0.6$~V and $T=60$~mK. (b) -- $P$ (left axis) and $\delta T_{eff}$ (right axis) as a function of $V_{DRIVE}$ for various values of $L$ (see legend). The detector QPC is defined with gate 2 (closed symbols) or gate 3 (open symbols). The drive QPC is placed about 40~$\mu$m upstream of the interaction region (not shown in Fig.~\ref{fig1}b). For this data, the electrostatic contribution was subtracted~\cite{electrostatics}. $\delta G_{DET}$ and $P$ are considerably smaller than in fig.~\ref{fig2}a, because of the hot EC cooling down on the way to the interaction region~\cite{eisenstein}. The data were taken at $V_C=-0.385$~V and $T=90$~mK.}\label{fig2}
\end{figure}

The linear response conductance of a QPC, $G=Tr\times e^2/h$, is proportional to its transparency $Tr$, the probability for an electron to be transmitted through the QPC. The energy dependence of $Tr$ can be used to convert a thermal gradient into an electric current~\cite{molenkamp,Dzurak_JPCM_1993,eisenstein}. Here we demonstrate a simpler and quantitative approach and use a QPC as a bolometer. In  leading order $\delta G_{DET}$ is proportional to the excess energy fluxes $\delta F_L$ and $\delta F_R$ impinging on it, respectively, from the left and right, and the second energy derivative of $Tr$:
\begin{equation}\label{eq_bolometer}
     \delta G_{DET}=e^2\frac{\partial^2Tr_{DET}}{\partial E^2}\frac{\delta F_R+\delta F_L}{2}.
\end{equation}
In the measurements discussed below we use either gate 2 or gate 3 (Fig.~\ref{fig1}b) to define our detector QPC at $Tr_{DET}\simeq0.1$.

In Fig.~\ref{fig2}a we present a typical bolometer measurement of $\delta G_{DET}$ versus  $V_{DRIVE}$. $\delta G_{DET}$ is parabolic at small $|V_{DRIVE}|$ and close to linear for $|V_{DRIVE}|\gtrsim100\mu$V. $\delta G_{DET}(V_{DRIVE})$ is  nearly symmetric in respect to the origin, indicating that the excess energy in the hot EC is independent of the sign of $V_{DRIVE}$. Similar results are obtained when the drive QPC is placed 40~$\mu$m upstream of the interaction region, see Fig.~\ref{fig2}b. We verified that our bolometer indeed probes the energy transfer rate ($P$) between the ECs within the interaction region. Corresponding experiments and the derivation of  formula~(\ref{eq_bolometer}) are presented in the Supplemental Material.

We calibrate the bolometer by measuring the $T$-dependence $\partial G_{DET}/\partial T$ in equilibrium and extracting $\partial^2Tr_{DET}/\partial E^2$. These two quantities are related via eq.~(\ref{eq_bolometer}) and a standard expression for the energy flux in equilibrium, $F_R=F_L=\pi^2(k_BT)^2/6h$~\cite{altimiras_natphys2010}. So obtained $P$ is given on the right/left axes in figs~\ref{fig2}a and~\ref{fig2}b, respectively. $P$ is in the fW range, meaning that just a tiny fraction ($\sim10^{-4}$) of the excess energy of the hot EC is absorbed in the cold EC. It is tempting to determine an effective excess temperature $\delta T_{eff}$ in the cold EC. For small changes one finds $\delta T_{eff}=2\delta G_{DET}/(\partial G_{DET}/\partial T)\ll T$. Here the factor of 2 accounts for the fact that in our experiment $\delta F_L=P$ and $\delta F_R=0$. $\delta T_{eff}$ is quantified in Figs.~\ref{fig2}a (bar) and~\ref{fig2}b (right axis).

 \begin{figure}[t]
 \begin{center}
  \includegraphics[width=\columnwidth]{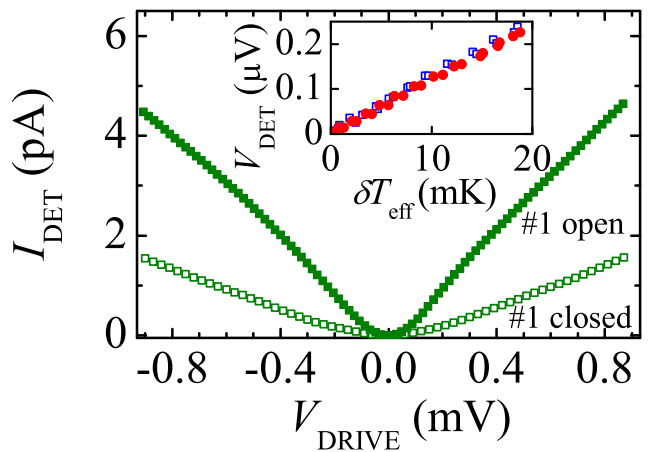}
   \end{center}
  \caption{Thermoelectric measurements.  $I_{DET}$ across the detector QPC 2 excited with the help of the drive QPC 8, $L=3\, \mu$m. The sign of $I_{DET}$ corresponds to injection of nonequilibrium electrons across the detector QPC, similar to the case of small magnetic fields~\cite{prokudina_PRB}. The effect is reduced when gate 1 is closed and the interaction region is isolated from the detector QPC, $L=0$ (Fig.~\ref{fig1}b). Inset: Proportionality of $V_{DET}$ and $\delta T_{eff}$ in the detector EC measured with the drive QPCs 5 (closed dots) and 6 (open squares) and  detector QPC 2, $L=5.2\, \mu$m. All the data corresponds to $V_C=-0.6$~V and $T=60$~mK.}\label{fig3}
\end{figure}

Non-zero $\delta T_{eff}$ in the cold EC generates a thermoelectric current ($I_{DET}$)  across the detector QPC. In Fig.~\ref{fig3}, $I_{DET}$ is plotted against $V_{DRIVE}$ for two choices of $L$.  This data closely resembles the bolometric response shown in Fig.~\ref{fig2}, which is a general feature of our measurements. The connection between the two
experiments becomes evident in the inset of Fig.~\ref{fig3}. The thermoelectric voltage, defined as $V_{DET}\equiv I_{DET}/G_{DET}$, is proportional to $\delta T_{eff}$ measured using the bolometer. The Seebeck coefficient of the detector QPC  $S=\delta V_{DET}/\delta T_{eff}\approx13\,\mu$V$/$K is comparable to previous measurements~\cite{molenkamp,Dzurak_JPCM_1993}. It is independent of the sign of $V_{DRIVE}$ and the choice of the drive QPC, as expected. The meaningful value of $S$ indicates that the cold EC is close to local thermal equilibrium and justifies our bolometric approach.

Observation of $P\propto V_{DRIVE}$ points at a breakdown of the momentum conservation for the energy exchange between counter-propagating ECs~\cite{prokudina_JETPL}. For a deeper analysis we use the kinetic equation approach \cite{gutmanPRB2009} and express $P$ in an inhomogeneous LL as
\begin{equation}\label{plasmon}
P=\frac{1}{h} \int\varepsilon f^{HOT}(\varepsilon) R_\varepsilon d\varepsilon\,,
\end{equation}
where $R_\varepsilon$ is the energy dependent backscattering probability of plasmons and  $f^{HOT}(\varepsilon)$ is their occupation number in the hot EC. In the limit of $|e V_{DRIVE}|\gg k_BT,\varepsilon$ we can assume  $f^{HOT}(\varepsilon)\propto |eV_{DRIVE}|/\varepsilon$. Hence $P\propto|V_{DRIVE}|$ indicates that backscattering is suppressed at high $\varepsilon$. Such a behavior is expected in a random disorder model with a finite correlation length $l_{corr}$ \cite{raikh}. The disorder potential absorbs momenta up to $\hbar l_{corr}^{-1}$ and enables transfer of energy quanta up to   $\varepsilon_0\sim\hbar ul_{corr}^{-1}$, where $u$ is the plasmon velocity~\cite{lunde}. With the magneto-plasmon velocity at $\nu=1$ estimated to be $u\sim10^7$cm/s~\cite{fujisawa} and with $\varepsilon_0\sim80\,\mu$eV determined from the onset of the linear slope of $P(V_{DRIVE})$ in Fig.~\ref{fig2}a, we find $l_{corr}\sim1{\text~}\mu$m for our device.

We gain more insights about plasmon scattering by studying $P$ in dependence on the ECs interaction length $L$. Using a fixed drive QPC (40~$\mu$m upstream of the interaction region) we vary $L$ in the range 0-6.3~$\mu$m by bending the hot EC with gates 6, 7 or 8 (see Fig.~\ref{fig1}b). As shown in Fig.~\ref{fig2}b $P$ stays constant as $L$ is increased between 2.2 and $6.3\,\mu$m. Obviously, this is inconsistent with random disorder scattering, for which $R_\varepsilon\propto L$~\cite{raikh}. Moreover, the heretical conclusion that part of the interaction region might be broken and would therefore not contribute to scattering is disproved in Fig.~\ref{fig3}. Instead, the independence of $P$ on $L$ indicates boundary scattering of plasmons at the entrance and exit of the interaction region as dominant energy transfer mechanism. As seen from fig.~\ref{fig2}b, $P$ depends on $L$ only for small $L\lesssim l_{corr}$, which can  be qualitatively explained by an overlap of the two boundaries and a sufficiently long-ranged interaction between the ECs (finite signal at $L=0$, see also fig.~\ref{fig3}).
\begin{figure}[t]
 \begin{center}
  \includegraphics[width=\columnwidth]{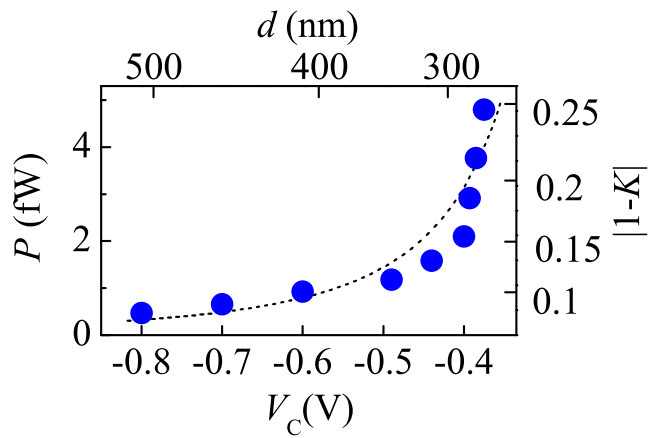}
   \end{center}
  \caption{Tuning the interaction. $P$ against $V_C$ (Fig.~\ref{fig1}b) for fixed $V_{DRIVE}=-0.4$~mV and $T=90$~mK. Gates 2/8 define the detector/drive QPCs, $L=3\, \mu$m. From the left to the right, the ECs separation $d$ reduces and $K$ is detuned from its noninteracting value 1 (see a scale on the right). $d$ obtained from a solution of the electrostatic problem is shown on the upper abscissae (see Supplemental Material for the details). The dashed line is a fit with parameters $v_F=1.2\times10^7$cm/s and $l_{bound}=770$nm, see text.}\label{fig4}
\end{figure}

The boundary scattering is related to the change of the plasmon velocity in the interaction region, where it is renormalized as $u=v_F/K$. Here $v_F$ is the Fermi velocity in the isolated EC and $K\ge1$ is a dimensionless LL interaction constant\cite{gutmanPRB2009}. Note that this process is a plasmon counterpart of a charge fractionalization at the LL boundary~\cite{safi1995}. At small energies the scattering obeys the Fresnel law $R_\varepsilon=\left[(1-K)/(1+K)\right]^2$. If $K$ varies smoothly across the length-scale $l_{bound}$, the reflection is suppressed for $\varepsilon\gtrsim\varepsilon_0=\hbar ul_{bound}^{-1}$, where $l_{bound}\sim1\mu$m replaces $l_{corr}$ considered above.

An independent indication for boundary scattering is the observed $P\propto V_{DRIVE}^2$ at weak driving  $|eV_{DRIVE}|\lesssim\varepsilon_0$, see Fig.~\ref{fig2}. This is expected for boundary scattering at $\varepsilon<\varepsilon_0$, as  $R_\varepsilon$ is constant in this case. In contrast, for disorder scattering~\cite{raikh} $R_\varepsilon\propto\varepsilon^2$ for  $\varepsilon<\varepsilon_0$, which would  result in $P\propto V_{DRIVE}^4$, similar to a perturbative calculation~\cite{prokudina_JETPL}.

The dashed line in Fig. \ref{fig2}a is a model curve based on Eq.~(\ref{plasmon}) assuming boundary scattering of plasmons. The only fit-parameters are  $\varepsilon_0=80\,\mu$eV, which sets the crossover from parabolic to linear $P(V_{DRIVE})$,  and  $K=1.12$. The interaction strength $|1-K|$ can be directly tuned by $V_C$. As shown in Fig.~\ref{fig4} for the case of $|eV_{DRIVE}|\gg\varepsilon_0$, $P$ sharply increases in the range $-0.8\,$V$<V_C<-0.4\,$V,  corresponding to $0.1<|1-K|\lesssim0.25$.

At our largest interaction $u\approx 0.75v_F$, which corresponds to a dimensionless LL conductance of $g\approx 0.5$. This is close to values reported in genuine 1DESs\cite{tunnel_yacoby,auslaender,bockrath}. We finally note that the electrostatic width of the central barrier is $d\simeq 300\,$nm (see upper axis of Fig.~\ref{fig4}), comparable to the depth of the 2DES and the width of the gate C. Our experiments are in the regime $d<l_{bound}$, for which the interaction is dominated by Coulomb coupling~\cite{prokudina_JETPL}. We obtain a reasonable agreement (dashed line in Fig.\ \ref{fig4}) evaluating the interaction as~\cite{giamarchi} $K=[1-(g_2/2\pi\hbar v_F)^2]^{-1/2}$, where $g_2=2e^2K_0(qd)/k$ is a matrix element of the Coulomb interaction at a wave vector $q=l_{bound}^{-1}$, $K_0$ is the Bessel function and $k\approx12.5$ is the dielectric constant~\cite{remark_perturb}. 

In summary, we studied the LL model out of thermal equilibrium based on counter-propagating quantum Hall ECs. The energy transfer between the ECs is consistent with elastic backscattering of collective density excitations at the boundaries of this hand-made LL. Counter-propagating quantum Hall ECs are a perfect candidate for refined tests of the LL theory, a first example being presented here.

We are grateful to I. Gornyi, I.S.~Burmistrov, V.T.~Dolgopolov and D.V.~Shovkun for discussions and to J.P.~Kotthaus for his input on early stages of this work. Financial support from RAS, RFBR, the Russian Ministry of Science and Education, as well, the German Excellence Initiative via the Nanosystems Initiative Munich (NIM) is acknowledged.


\begin{thebibliography}{50}
\bibitem{electrostatics} The trivial electrostatic contribution to $G_{DET}$ arises from the change of the electron density in the biased drive circuit owing to the in-plain gating effect. For $|\delta G_{DET}/G_{DET}|\ll1$ it is asymmetric in drive bias $\delta G_{DET}\propto V_{DRIVE}$. This contribution was independently calibrated in experiments with an open drive QPC, when the bolometric contribution is absent.
\bibitem{remark_perturb} At $|K-1|\ll1$ a renormalization of the plasmon velocity is small, and we expect a perturbative description to be adequate as well. Unlike in Ref.~\cite{prokudina_JETPL}, in this case one has to account for a pairwise electron-electron scattering at the boundaries of the interaction region.
\end{thebibliography}
\end{document}


\title{Tunable non-equilibrium Luttinger liquid based on counter-propagating edge channels. Supplemental Material.}
\author{M.G.~Prokudina}
\affiliation{Institute of Solid State Physics, Russian Academy of
Sciences, 142432 Chernogolovka, Russian Federation}
\author{S.~Ludwig}
\affiliation{Center for NanoScience and Fakult\"{a}t f\"{u}r Physik,
Ludwig-Maximilians-Universit$\ddot{\text{a}}$t,
Geschwister-Scholl-Platz 1, D-80539 M$\ddot{\text{u}}$nchen,
Germany}
\author{V. Pellegrini}
\affiliation{NEST, Istituto Nanoscienze-CNR and Scuola Normale Superiore, Piazza San Silvestro 12, I-56127 Pisa, Italy}
\author{L. Sorba}
\affiliation{NEST, Istituto Nanoscienze-CNR and Scuola Normale Superiore, Piazza San Silvestro 12, I-56127 Pisa, Italy}
\author{G.~Biasiol}
\affiliation{CNR-IOM, Laboratorio TASC, Area Science Park, I-34149 Trieste, Italy }
\author{V.S.~Khrapai}
\affiliation{Institute of Solid State Physics, Russian Academy of
Sciences, 142432 Chernogolovka, Russian Federation}

\maketitle

\section{Bolometric response of a QPC}

Here we derive an expression for a linear response conductance of a quantum point contact (QPC) in a weakly non-equilibrium non-interacting spinless 1D electron system (1DES). The (nonequilibrium) distribution functions of the carriers  in the right/left edge channels incident on the QPC are denoted, respectively, as $f_R$ and $f_L$. The corresponding chemical potentials are determined from the conservation of the particle number:
\begin{equation}\label{mu}
    \mu_{R,L}\equiv\int_0^\infty f_{R,L}(E) dE=\mu_0\pm eV/2,
\end{equation}
where $E$ is the energy, $\mu_0$ is the chemical potential at equilibrium, $e$ is the elementary charge and $V$ is the bias voltage across the QPC. The dispersion relation is linearized near the Fermi surface, which gives rise to the energy independent density of states. The net current through the QPC is determined by the energy-dependent QPC transparency $Tr(E)$:
\begin{equation}\label{current}
    I=\frac{e}{h}\int_0^\infty Tr(E) (f_R-f_L)dE,
\end{equation}
which can be differentiated in respect to $V$ to give the conductance:
\begin{eqnarray}
G\equiv\frac{\partial I}{\partial V}=\frac{e}{h}\int_0^\infty Tr(E) \left(\frac{\partial f_R}{\partial V}-\frac{\partial f_L}{\partial V}\right)dE= \nonumber \\
     =\frac{-e^2}{2h}\int_0^\infty  Tr(E) \left(\frac{\partial f_R}{\partial E}+\frac{\partial f_L}{\partial E}\right) dE, \label{cond}
\end{eqnarray}
where we used the linear response relation $\partial f_{R,L}/\partial V=\pm e/2\partial f_{R,L}/\partial E|_{V=0}$, which follows from the fact that the bias voltage doesn't affect the distribution functions apart from shifting $\mu_R$ and $\mu_L$. Eq.~(\ref{cond}) simplifies close to equilibrium, where the distributions $f_{R,L}$ differ from 0/1 only within a narrow energy window. In this case the transparency $Tr$ is almost constant and we account for its energy-dependence up to the second order $Tr=Tr^0+Tr'\epsilon+Tr''\varepsilon^2/2$. Here, $\varepsilon\equiv E-\mu_0$  and $Tr^0,Tr',Tr''$ are, respectively, the transparency and its first and second derivatives at $E=\mu_0$. In these notations eq.~(\ref{cond}) reduces to:
\begin{equation}
    G=G_0+G_1+G_2 \label{cond_integrals}
\end{equation}
where we used three identities:
\begin{equation*}
G_0=-\frac{e^2}{h}Tr^0\int_{-\mu_0}^\infty\frac{\partial \overline{f}}{\partial\varepsilon} d\varepsilon=\frac{e^2}{h}Tr^0
\end{equation*}
\begin{equation*}
G_1=-\frac{e^2}{h}Tr'\int_{-\mu_0}^\infty\varepsilon\frac{\partial \overline{f}}{\partial\varepsilon} d\varepsilon=0
\end{equation*}
\begin{eqnarray}
G_2=-\frac{e^2}{2h}Tr''\int_{-\mu_0}^\infty\varepsilon^2\frac{\partial \overline{f}}{\partial\varepsilon} d\varepsilon= \nonumber \\
=\frac{e^2}{h}Tr''\left(\int\varepsilon \overline{f}d\varepsilon-\mu_0^2/2\right)=e^2Tr''(\overline{F}-F_0), \nonumber
\end{eqnarray}
\begin{equation*}
\text{with\,}  \overline{f}=\frac{f_L+f_R}{2} \text{\,and\,}  \overline{F}=\frac{F_L+F_R}{2}
\end{equation*}
which follow from the properties of the Fermi distribution function and eq.~(\ref{mu}). Here $F_R$ and $F_L$ is a total energy flux in the two edge channels (ECs) and $F_0$ is its value at a zero temperature ($T=0$). Note that in equilibrium $F^{eq}_{R,L}=\pi^2(k_BT)^2/6h$ so that the term $G_2$ in eq.~(\ref{cond_integrals}) accounts also for the $T$-dependence of $G$. Summarizing, we find for a deviation of the conductance caused by heating of the ECs incident on the QPC:
\begin{equation}\label{eq_bolometer}
     \delta G=\delta G_2=e^2Tr''\times\frac{\delta F_R+\delta F_L}{2},
\end{equation}
where $\delta F_{R,L}$ are the excess energy fluxes carried by corresponding ECs in respect to $F^{eq}_{R,L}$ at a given $T$.
Similarly, at $V=0$ one obtains in the first order in $\varepsilon$ an expression for thermoelectric current  from eq.~(\ref{current}):
\begin{equation}\label{thermo}
    I_{therm}=eTr'\times\left(\delta F_R-\delta F_L\right).
\end{equation}
Equations~(\ref{eq_bolometer}) and~(\ref{thermo}) express the bolometric and the thermoelectric responses of a QPC out of equilibrium in terms of the energy dependence of its transparency. As explained in the main paper, from these expressions one can calibrate the bolometric response via a conductance temperature dependence and, likewise, evaluate a Seebeck coefficient (thermopower). The latter is defined as  $S=V_{therm}/\delta T_{eff}$, where $V_{therm}\equiv I_{therm}G^{-1}$ is the thermoelectric voltage and $\delta T_{eff}\equiv3h(\delta F_R-\delta F_L)/(\pi^2k_B^2T)\ll T$ is the effective temperature gradient.

\begin{figure}[t]
 \begin{center}
  \includegraphics[width=0.8\columnwidth]{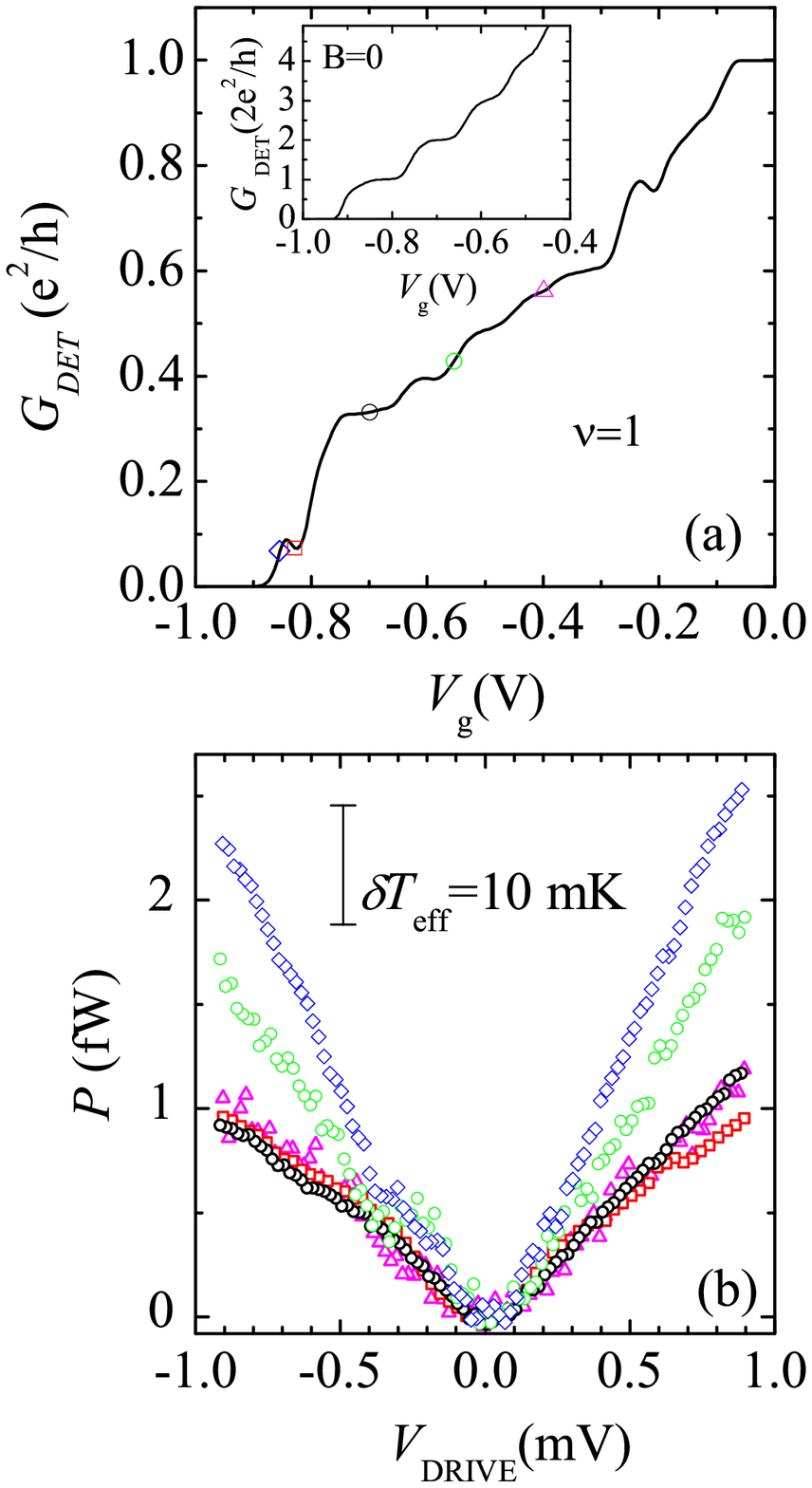}
   \end{center}
  \caption{Varying the transparency of the detector QPC. (a) -- A typical gate voltage dependence of the detector-QPC conductance at $nu=1$ (body) and in zero magnetic field (inset) with a series resistance subtracted. (b) -- Measured energy transfer rate $P$ as a function of the excitation bias in the hot EC. Different symbols correspond to different transparencies in the range $0.06<Tr^0<0.6$, see the data points in (a). The data are taken for a detector QPC 2 and drive QPC 8, see fig.~1b of the main paper. The scale of $P$ corresponding to the
  excess temperature of 10~mK is given by the vertical bar.
  }\label{fig_Teff}
\end{figure}

The nearly perfect proportionality $V_{thermo}\propto\delta T_{eff}$  observed in fig.~3 of the main paper is in agreement with Eqs.~(\ref{eq_bolometer})and~(\ref{thermo}) provided $\delta F_L\equiv0$, i.e. when one of the ECs remains at equilibrium. Yet the evaluated energy tarnsfer rate $P=\delta F_R$ depends on the detector QPC transparency. At moderate excitations $|V_{DRIVE}|\leq0.5$mV, see Fig.~\ref{fig_Teff}b, the $P$ varies by at most a factor of $\sim2$ in the range $0.06<Tr^0<0.6$ (see the data points in Fig.~\ref{fig_Teff}a). This uncertainty is still acceptable in light of the vast variation of the bolometric sensitivity by a factor of $\sim20$ for the same data. Summarizing the above, the lowest order approximations provide a consistent description of the experiment and permit a reliable estimate of the excess energy flux in the detector EC.

Note, that in our derivation we assumed that the width of the nonequilibrium distribution is small compared to the characteristic scale $\Omega$ of the energy dependence $Tr(E)$ in the detector QPC. This is straightforward to verify. Close to pinch-off, the dependence is close to the exponential~\cite{buettiker1990} $Tr\propto exp(-E/\Omega)$, so that $Tr'\approx Tr^0/\Omega$. In our experiment, $S\approx 13\mu V/K$ (see the inset of fig.~3 of the main paper), which gives an estimate $\Omega=\pi^2k_B^2T/3eS\sim100\mu$eV from eq.~(\ref{thermo}). On one hand $\Omega\gg T,\delta T_{eff}$, while on the other hand $\Omega$ is the same order of magnitude as the bandwidth $\varepsilon_0\approx80\mu$eV of the energy relaxation (see the main paper). That is, the derivation would perfectly hold if the electrons in the cold EC were at a local equilibrium with an effective excess temperature $\delta T_{eff}\sim50$~mK (scale bar in Fig.~\ref{fig_Teff}b). And break down in the opposite case. Apparently, the experimental results point to a (partial) carrier thermalization, which is not surprising in view of strong dephasing in Fabri-Perot~\cite{vanderWiel_PRB2003} and  $\nu=1$ Mach-Zender interferometers~\cite{heiblum_MZ_2003} at small excitation energies.

\section{Proof of the inter-EC energy transfer}

\begin{figure}[t]
 \begin{center}
  \includegraphics[width=0.8\columnwidth]{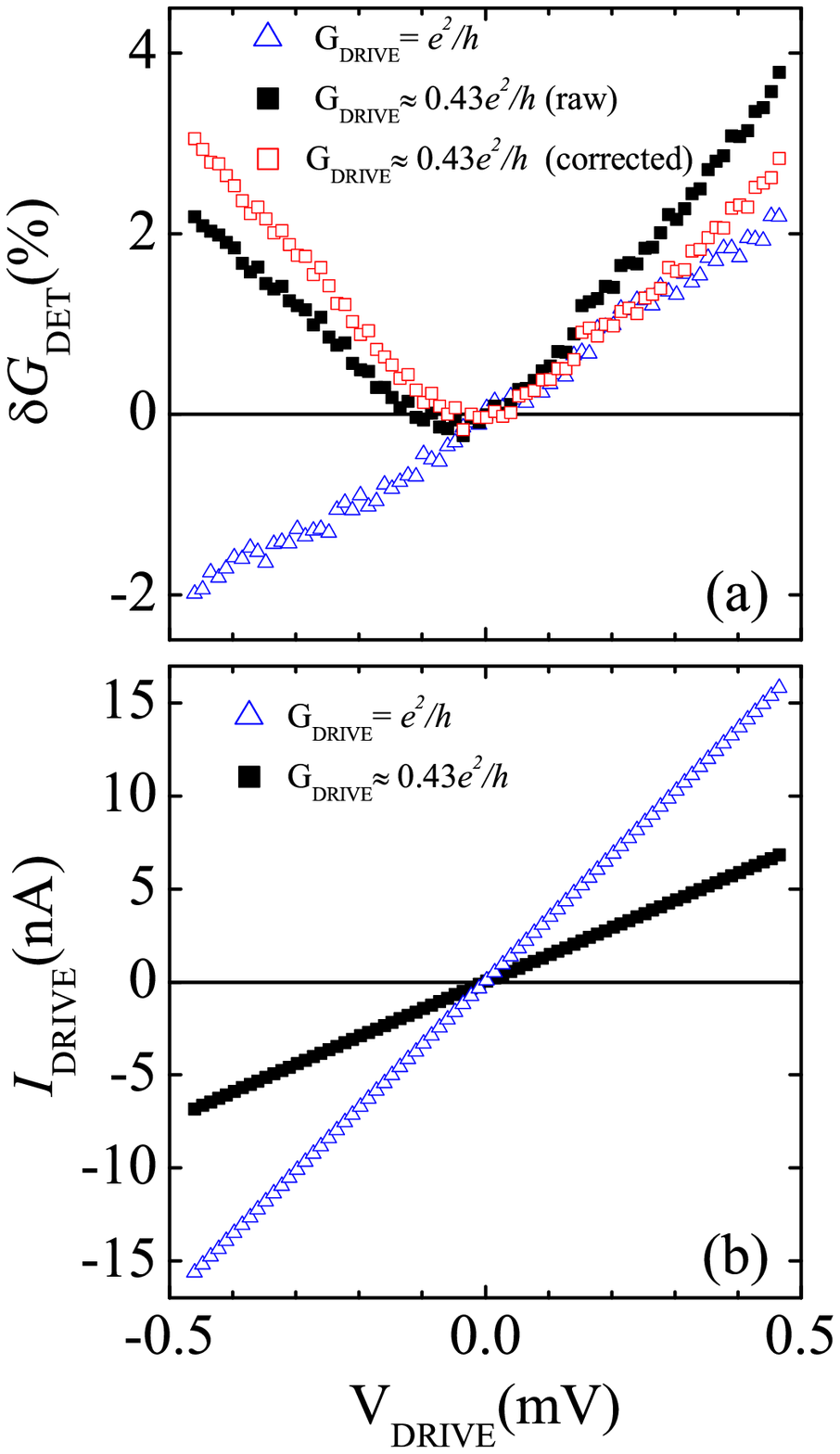}
   \end{center}
  \caption{Electrostatic contribution to the detector conductance. (a) -- Measured variation of the detector conductance as a function of the excitation bias in the drive circuit in case of open (blue triangles) and partially transparent (black squares) drive-QPC.  The latter data after correction for the electrostatic contribution are shown by red squares.  The detector is defined with gate 2, see fig.~1b of the main paper, and the sketch of the experiment is depicted in fig.~\ref{chirality1}c2. (b) -- The $I$-$V$ characteristics of the drive-QPC measured simultaneously with the data of (a). }\label{statics}
\end{figure}

\begin{figure}[t]
 \begin{center}
  \includegraphics[width=\columnwidth]{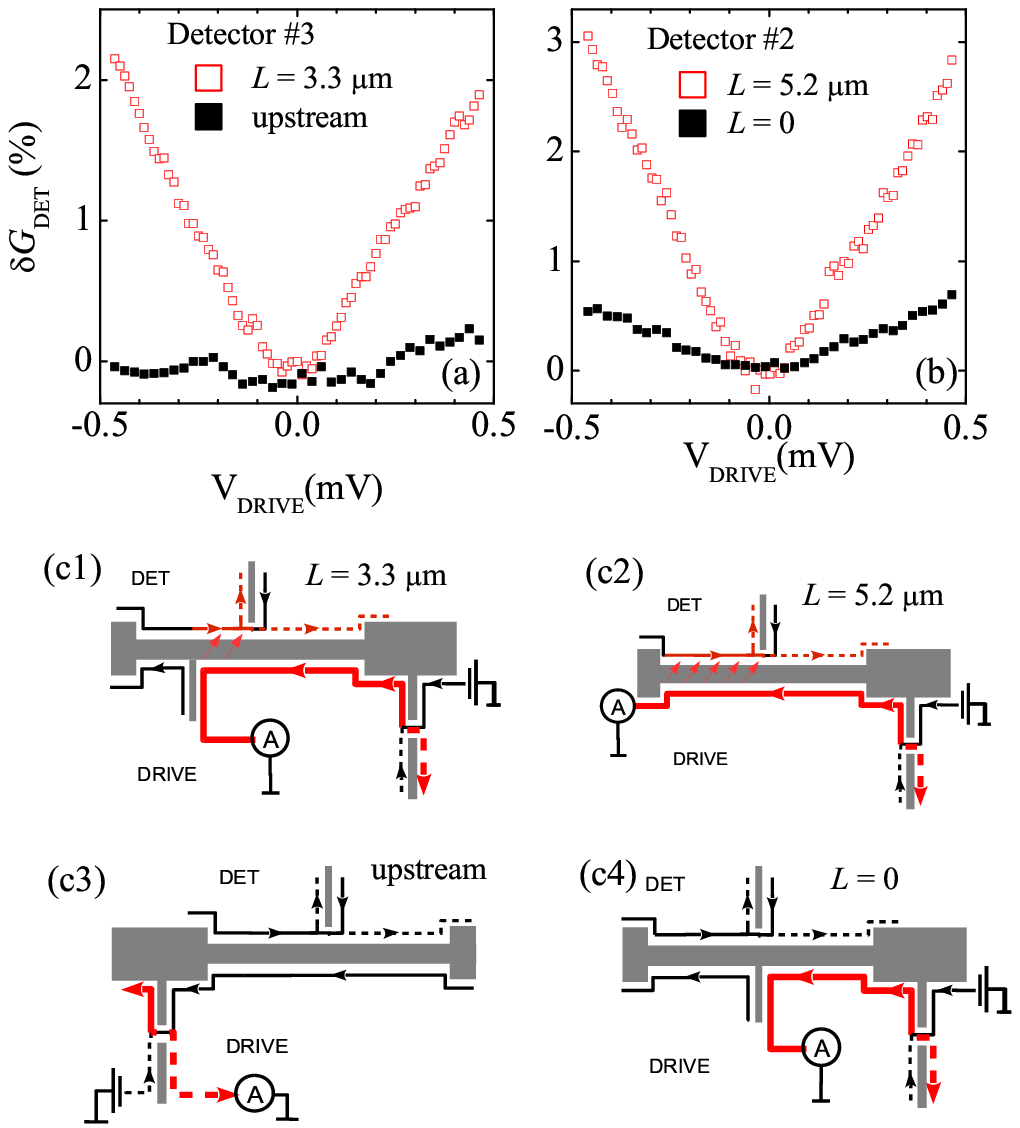}
   \end{center}
  \caption{Verification of the origin of the bolometric signal. (a), (b) -- Change of the detector conductance as a function of the excitation bias after accounting for the electrostatic contribution. The legends correspond to the experimental schemes depicted in (c1), (c2), (c3) and (c4). The values of the interaction length $L$ are given in the legends.}\label{chirality1}
\end{figure}

In this section we present test experiments that prove inter-EC energy transfer as the origin of the bolometric signal in the detector-QPC. First of all we discriminate the bolometric signal from a spurious electrostatic coupling effect. The experimental scheme is depicted in fig.~\ref{chirality1}c2. We measure the change of the detector conductance $\delta G_{DET}$ as a function of the bias $V_{DRIVE}$ applied in the drive circuit. For a partially transparent drive-QPC (black squares in fig.~\ref{statics}a), the signal $\delta G_{DET}>0$  is a factor of 2 asymmetric in respect to bias reversal and corresponds to a temperature increase. For a fully open drive-QPC (blue triangle in fig.~\ref{statics}a), however, the signal is fully antisymmetric, which is a result of electrostatic coupling between the detector-QPC constriction and hot-EC in the drive circuit (gating). Within the linear approximation $\delta G_{DET}\propto\phi$, where  $\phi$ is the electrostatic potential of the hot-EC. It's straightforward to show that $\phi=I_{DRIVE}\times(h/e^2+R_{cont})$, where $I_{DRIVE}$ is the current measured in the drive-circuit and $R_{cont}\sim1\,k\Omega$ is the resistance of the ohmic contact which connects the hot-EC and the $I-V$ converter (see the sketch of fig.~\ref{chirality1}c2). The dependencies  $I_{DRIVE}$ vs $V_{DRIVE}$ measured in an open and partially transparent drive-QPC are plotted fig.~\ref{statics}b as blue triangles and black squares, respectively. The slope ratio is $\approx2.3$ which allows to correct the bolometric data of fig.~\ref{statics}a for the gating effect (both effects are small, hence additive). As a result, the asymmetric curve (black squares) is transformed into the almost symmetric one (red squares). Such a procedure to subtract the gating contribution was performed below where necessary.

The following experiments verify that the bolometric signal comes from the interaction between the ECs counter propagating along the narrow segment of the central gate (interaction region). As seen from fig.~\ref{chirality1}a, when we choose the drive-QPC such that the interaction region is upstream of it (fig.~\ref{chirality1}c3), no detector response is observed (black squares). This is a result of chirality of the heat propagation in quantum Hall regime~\cite{eisenstein}. Alternatively, one can suppress the bolometric signal by reducing the interaction length to $L=0$ via a proper gating, see black squares in fig.~\ref{chirality1}b and a sketch (c4). We believe, that a residual signal in this case is a result of long range Coulomb interaction between the cold-EC and the hot-EC, see also fig.~3 of the main paper. On the other hand, the full bolometric signal is restored when the hot-EC and the cold-EC are allowed to interact over a few microns interaction region, see red squares in figs.~\ref{chirality1}a and~\ref{chirality1}b and, respectively, the sketches (c1) and (c2). This is a clear demonstration that the concept of nonequilibrium interaction between the counterpropagating ECs is fully consistent with our experiment.

\section{Energy relaxation on plasmon language}

Energy transfer between two counterpropagating ECs in quantum Hall regime at $\nu=1$ is convenient to describe within a concept of spinless Luttinger liquid (Ll). As discussed in the main paper, the elementary excitations in such a system are bosonic density excitation --- plasmons. Following Ref.~\cite{gutmanPRL2008}, consider a finite Ll connected to semi-infinite Fermi-liquids on both sides. This geometry is analogous to a system of counter-propagating ECs with a nonzero inter-EC interaction within the interaction region.
Outside the interaction region the plasmon distribution function is conserved.  The incoming plasmon distributions $B_R^{in},B_L^{in}$ are obtained, respectively, from the right/left moving particle-hole distribution in the Fermi liquid leads~\cite{gutmanPRL2008}:
\begin{equation*}
B_{R,L}^{in}(\varepsilon)=\frac{1}{\varepsilon}\int n(E)\left[2-n(E-\varepsilon)-n(E+\varepsilon)\right]dE,
\end{equation*}
where $\varepsilon$ is the plasmon energy and $n(E)$ is the incoming distribution function of the right/left moving electrons with the energy $E$. This expression is a sum of the form-factors for creation and annihilation of electron-hole pairs. Note, that the eq.~(5) of Ref.~\cite{gutmanPRL2008} is different by a factor of 2, which we believe to be a misprint. In our experiment, the nonequilibrium (double-step) electronic distribution is created by a QPC in one (say, right moving) EC. As follows from the above equation $B_R^{in}(\varepsilon)=1+2Tr(1-Tr)(eV_{DRIVE}/\varepsilon-1)$, where $V_{DRIVE}$ is the drive bias applied across the drive-QPC, $Tr$ is its transparency and $\varepsilon\leq eV_{DRIVE}$. The equilibrium Fermi distribution in the other (left moving) EC corresponds to $B_L^{in}(\varepsilon)=1+2f_B(\varepsilon)$, where $f_B(\varepsilon)$ is the equilibrium Bose-Einstein distribution  with a base temperature $T\approx60$mK. At relevant plasmon energies $\varepsilon_0\sim50\mu$eV, $f_B(\varepsilon)\ll1$, i.e. we can safely use the zero-$T$ approximation $B_L^{in}=1$. In the Ll the plasmon distributions are modified owing to a plasmon backscattering  at the boundaries of the interaction region (see the main paper). The distribution of the outcoming left moving plasmons is increased by $\delta B_L^{out}=R_\varepsilon (B_R^{in}-B_L^{in})$, where $R_\varepsilon$ is the energy-dependent scattering probability. Hence, we get for the inter-EC energy transfer rate:
\begin{equation*}
P=\frac{1}{2h}\int \delta B_L^{out}(\varepsilon)\varepsilon d\varepsilon=
\end{equation*}
\begin{equation}\label{relaxation}
=\frac{Tr(1-Tr)}{h}\int_0^{eV} R_\varepsilon (eV-\varepsilon)d\varepsilon.
\end{equation}

\begin{figure}[t]
 \begin{center}
  \includegraphics[width=0.8\columnwidth]{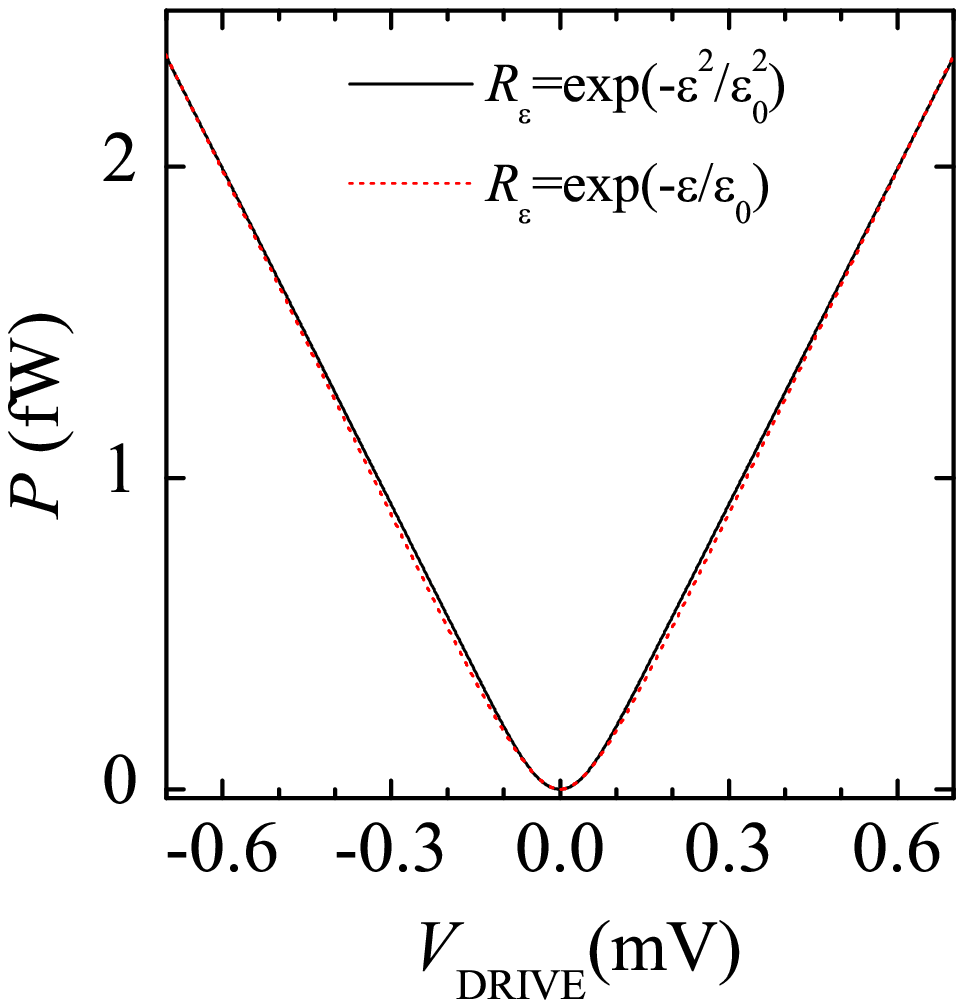}
   \end{center}
  \caption{Different shapes of the high-energy cutoff of the plasmon scattering probability. Using eq.~(\ref{relaxation}) we compare the energy transfer rates calculated for different energy dependencies of the plasmon scattering probability $R_\varepsilon$, see legend. Nearly the same results are obtained for gaussian and simple exponential dependencies, respectively, with parameters $K\approx1.119$, $\varepsilon_0=80\mu$eV and $K\approx1.132$, $\varepsilon_0=60\mu$eV. The former curve corresponds to the best fit of the experiment in fig.~2a of the main paper.}\label{fitR}
\end{figure}

Eq.~(\ref{relaxation}) allows to express the energy relaxation between the counter propagating ECs in terms of the (small) plasmon backscattering probability $R_\varepsilon\ll1$. In our fits we assumed a modified Fresnel law $R_\varepsilon=(1-K)^2/(1+K^2)\times F(\varepsilon)$, where $K$ is the interaction constant defined below and  the {\it ad hoc} factor $F(\varepsilon)$ accounts for a suppressed backscattering of high-energy plasmons owing to a finite length-scale of the inhomogeneity at the boundaries of the Ll.
Little is known about $F$ and we assume two different exponential dependencies $F=\exp(-\varepsilon^2/\varepsilon_0^2)$
and $F=\exp(-\varepsilon/\varepsilon_0)$ in the following. The calculated energy transfer rate $P(V_{DRIVE})$ is plotted
in fig.~\ref{fitR}. For both choices of $F(\varepsilon)$ a crossover from parabolic to linear dependence at increasing $V_{DRIVE}$ is observed.
Moreover, for a proper choice of parameters $K,\varepsilon_0$ the results are almost indistinguishable.
This allows to roughly estimate the uncertainties of our fit parameters as 10\% in $K$ and 30\% in $\varepsilon_0$.

\section{Interaction constant \textit{K}}

The value of the interaction constant $K$ can be determined from the matrix element of the Coulomb interaction, see Ref.~\cite{giamarchi}:
\begin{equation*}
u=v_F\left[(1+y_4)^2-(y_2)^2\right]^{1/2},
\end{equation*}
where $u$ is the plasmon velocity, $v_F$ is the Fermi velocity in the absence of interactions and $y_{i}=g_i/(hv_F)$ are
 the dimensionless matrix elements of the intra-EC ($i=4$) and inter-EC ($i=2$) Coulomb interaction. In our experiment
  $g_2\neq0$ only within the interaction region, whereas a much stronger intra-EC interaction $g_4\gg g_2$ can be assumed
   constant everywhere. Hence, we can rewrite this equation in terms of a renormalized Fermi velocity:
\begin{equation}\label{K}
u=v^*_F\left[1-(y^*_2)^2\right]^{1/2},
\end{equation}
where $v^*_F=v_F+g_4/h$ is the interaction-renormalized Fermi velocity in the Fermi liquid leads 
and $y^*_2=g_2/(hv^*_F)$ is the renormalized dimensionless inter-EC interaction in the Ll. 
Importantly,  $v^*_F\gg v_F$ is nothing but a magnetoplasmon velocity of an isolated EC, 
whereas $u<v^*_F$ is the magnetoplasmon velocity in the interaction region. 
Hence, it's their ratio $K=v^*_F/u$ that enters the Fresnel law and defines the 
plasmon scattering at the boundaries of the interaction region, see above. 
Note that, as follows from eq.~(\ref{K}), the lowest order correction to $K$ is second order in interaction, 
which is different from the case $g_4=g_2$ considered, e.g., in Ref.~\cite{gutmanPRB2009}.

The matrix element $g_2$ for a given inter-EC separation $d$ can be evaluated in two ways: At zero momentum $q=0$, one has to introduce a screening radius $r$ of the Coulomb interaction for convergence:
\begin{equation}\label{qzero}
  g_2=\int_{-r}^{+r}\frac{e^2}{k(x^2+d^2)}dx=\frac{2e^2}{k}{\rm asinh}(r/d),
\end{equation}
where $k=12.5$ is the dielectric constant of GaAs. Alternatively, one can neglect screening but evaluate the matrix element at a finite momentum corresponding to a relevant length-scale $q=1/l_{corr}$:
\begin{equation}\label{qfinite}
  g_2=\int_{-\infty}^{+\infty}\frac{e^2\exp(-iqx)}{k(x^2+d^2)}dx=\frac{2e^2}{k}K_0(qd),
\end{equation}
where $K_0$ is the modified Bessel function of the second kind. In practice, the best fits to the experimental data obtained with eqs.~(\ref{qzero}) and (\ref{qfinite}) are almost indistinguishable provided $q\approx(2r)^{-1}$. This is a result of logarithmic behavior $g_2\propto\log(2r/d)$ and $g_2\propto\log(qd)$ at small $d$. The fit in fig.~4 of the main paper was performed for the bare Coulomb potential with the help of eqs.~(\ref{relaxation}),(\ref{K}) and (\ref{qfinite}). The value of the magnetoplasmon velocity  was chosen in the range $u\sim10^7$cm/s as we expect for a soft edge at $\nu=1$ (based, e.g., on a recent data for $\nu=2$~\cite{fujisawa}). In turn, the value of the correlation length $l_{corr}$ is constrained by the bandwidth $\varepsilon_0\approx 80\mu$eV. The best fit corresponds to $\hbar u/l_{corr}\approx100\mu$eV, which is  reasonably close to the experiment.

\section{Edge channels separation}
\begin{figure}[t]
 \begin{center}
  \includegraphics[width=0.8\columnwidth]{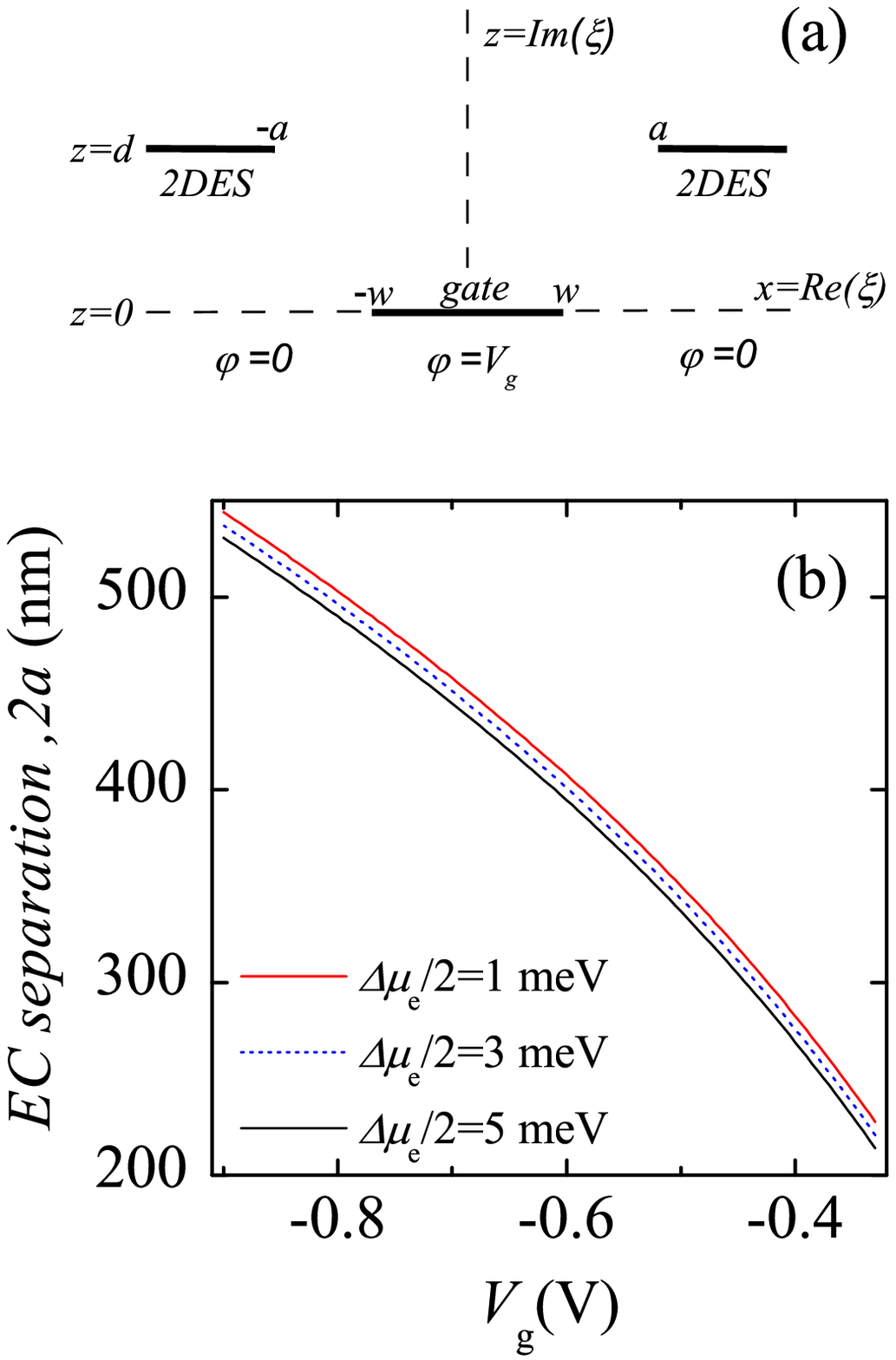}
   \end{center}
  \caption{Numeric calculations of the separation between the stripe-gate defined 2DES edges. (a) -- conformal mapping used to calculate the bare electrostatic potential created by the stripe-gate in the 2DES plain. (b) -- gate voltage dependence of the separation $2a$ between the ECs for 2DES depth of $d=200$~nm, gate width of $2w=200$~nm and several values of the spin gap $\Delta\mu_e$ at $\nu=1$.}\label{fitdistance}
\end{figure}

The strength of the inter-EC Coulomb interaction is determined by the distance $2a$ between the counterpropagating edges, tunable by the voltage $V_g$ on the central gate, see fig.~1b of the main paper. We evaluate this with the help of a simplified analytic solution. First we use a conformal mapping approach~\cite{larkin} to find the potential $\varphi_g$ a stripe gate creates in the 2DES plain. Here we assume an infinite metallic gate of width $2w$  (with a potential $V_g$) on a so-called pinned surface, i.e. the electrostatic potential of the remainder of the surface is fixed at $\varphi=0$. We chose this boundary condition for a much simpler solution it gives. The conformal mapping is straightforward $\xi=x+iz$ (see fig.~\ref{fitdistance}a) and we obtain~\cite{larkin}:
\begin{equation*}
  \varphi_g(x)=\frac{V_g}{\pi}Im[-\ln(x+w+id)+\ln(x-w+id)]=
\end{equation*}
\begin{equation}\label{conformal}
 =\frac{V_g}{\pi}\left[-\arctan\left(\frac{d}{x+w}\right)+\arctan\left(\frac{d}{x-w}\right)\right],
\end{equation}
where $d$ is the depth of the 2DES below the surface and ${\arctan\in(0:\pi)}$. The potential (\ref{conformal}) is just the bare potential in the absence of the 2DES. Next we add two semi-infinite electron layers ($|x|>a$) with a fixed electrons density $n_S$, for the 2DES is in the incompressible state at $\nu=1$. Note that in order to keep the boundary conditions satisfied one also has to introduce image charges at $z=-d$ and account for their potential. The potential $\varphi_e$ created by the electron layer and image charges is easily found. For example at the edge of the 2DES $x=a,z=d$:
\begin{equation}\label{2DESpotential}
  \varphi_e(a)/\varphi_0=1-\frac{1}{\pi}\arctan\left(\frac{a}{d}\right)-\frac{a}{2\pi d}\ln\left(1+\frac{d^2}{a^2}\right),
\end{equation}
where $\varphi_0=4\pi en_Sd/k\approx-0.27$~V is the 2DES potential at infinity ($|x|\rightarrow\infty$). The total electrostatic potential is given by the sum of $\varphi_e+\varphi_g$. The difference in potential energies of an electron at the gate-defined edge and at infinity is given by:
\begin{equation}\label{edgeenergy}
  dE(a)= e[\varphi_g(a)+\varphi_e(a)-\varphi_0].
\end{equation}
At $\nu=1$ the same energy difference equals half the chemical potential jump (interaction enhanced spin-gap) across the spin-gap between the Landau levels $dE(a)=\Delta\mu_e/2$. For a given $Vg$ and $\Delta\mu_e$ eqs.~(\ref{conformal}), (\ref{2DESpotential}) and (\ref{edgeenergy}) are satisfied for certain $a$, which defines the distance between the counter-propagating edges. Results of such calculations are shown in fig.~\ref{fitdistance}b for several values of $\Delta\mu_e$. Note that the actual value of the enhanced spin-gap in GaAs is not known accurately~\cite{nicholas,aristov}. Nevertheless even a huge variation in $\Delta\mu_e$ gives rise only to minor uncertainties in $a$, see fig.~\ref{fitdistance}b. This is a result of strong gradient of the electrostatic potential (in-plain electric field) created by the gate near the 2DES edge. The simulation in the fig.~4 of the main paper has been performed for $\Delta\mu_e=0$.

No doubt that our approach to calculate the EC separation is rather simplified. First, the pinned surface boundary conditions is not the case at low temperatures~\cite{larkin}. Second, the screening of the external potential results in formation of compressible strip at the edge of the 2DES. Accounting for these effects requires  much more involved approaches and might improve the agreement between the experiment and simulations in fig.~4 of the main paper. Yet, it is a-priori clear that the distance between the ECs in our structure is in a few 100~nm range. Hence, the absolute value of the evaluated dimensionless interaction $g_2$ in our quantum Hall based Ll is not expected to change appreciably.